\documentclass[reqno,12pt]{amsart}
\usepackage{amssymb, latexsym, euscript}

\newtheorem{remark}{\bfseries{Remark}}[section]

\newtheorem{theorem}{Theorem}[section]
\newtheorem{proposition}[theorem]{Proposition}

\newtheorem{lemma}[theorem]{Lemma}
\newtheorem{definition}[theorem]{Definition}

\numberwithin{equation}{section}
\newcommand{\kerr}{\mbox{Ker}}

\begin{document}
\title[Lax-Phillips scattering theory...]{Lax-Phillips scattering theory for $\mathcal{PT}$-symmetric $\rho$-perturbed operators}
\author[P.~Cojuhari]{Petru A.~Cojuhari }
\author[S.~Kuzhel]{Sergii~Kuzhel}

\address{AGH University of Science and Technology \\ Department of Applied Mathematics \\
30-059 Krakow, Poland} \email{cojuhari@agh.edu.pl}

\address{AGH University of Science and Technology \\ Department of Applied Mathematics \\
30-059 Krakow, Poland} \email{kuzhel@mat.agh.edu.pl}

\keywords{$\mathcal{PT}$-symmetric operators, Lax-Phillips scattering theory, $S$-matrix, operator of $\mathcal{C}$-symmetry.}

\subjclass[2000]{Primary 47A55, 47B25; Secondary 47A57, 81Q15}
\maketitle
\begin{abstract}
The $S$-matrices corresponding to $\mathcal{PT}$-symmetric $\rho$-perturbed operators
are defined and calculated by means of an approach based on an operator-theoretical interpretation of the Lax-Phillips scattering theory.
\end{abstract}
\section{Introduction}
The development of $\mathcal{P}\mathcal{T}$-symmetric quantum mechanics (PTQM) attracts a lot of interests during the past decade
\cite{BE, MO}. The PTQM is based on the idea that the Hermiticity condition, which is stated as an axiom of quantum mechanics,
may be replaced by a certain less mathematical but more physical condition of symmetry without losing any of the essential physical features of
quantum mechanics.

The Hamiltonians of PTQM are not self-adjoint with respect to the initial Hilbert space's inner product and their
`physical symmetries' do not depend on the choice of inner product. One of typical examples is the non-selfadjoint Hamiltonian
$$
    H=-\frac{d^2}{dx^2} + x^2(ix)^\epsilon, \qquad  0\leq\epsilon<2
$$
in $L_2(\mathbb{R})$. The spectrum of $H$ is real and positive \cite{B1,D2} and $H$ has
the property of  $\mathcal{P}\mathcal{T}$-symmetry $
{\mathcal P}{\mathcal
T}H=H{\mathcal P}{\mathcal T},$
where the space reflection
(parity) operator $\mathcal{P}$ and the complex conjugation operator
$\mathcal{T}$ are defined as
$({\mathcal P}f)(x)=f(-x)$ and $({\mathcal T}f)(x)=\overline{f(x)}.$

The property of $\mathcal{PT}$-symmetry depends  on the choice of operators $\mathcal{P}$ and $\mathcal{T}$, which can be different for
various underlying Hilbert space $\mathfrak{H}$ and various non-selfadjoint operators $H$.
This means that Hamiltonians of PTQM may have the property of $\mathcal{PT}$-symmetry realized by different operators
$\mathcal{P}$ and $\mathcal{T}$.

However, the linear operator $\mathcal{P}$ is always a unitary involution in $\mathfrak{H}$, that is
$\mathcal{P}^2=I$, $(\mathcal{P}f,\mathcal{P}g)=(f,g)$; and the anti-linear operator $\mathcal{T}$ is a conjugation operator in $\mathfrak{H}$,
that is $\mathcal{T}^2=I$, $(\mathcal{T}f,\mathcal{T}g)=(f,g)$.
This observation allows one to develop the theory of $\mathcal{PT}$-symmetric operators in some abstract setting
keeping in mind the properties of operators $\mathcal{P}$ and $\mathcal{T}$ mentioned above (subsection 3.1).

Nowadays, scattering problems related to the Schr\"{o}dinger-type
differential expression
\begin{equation}\label{gere1}
l(\cdot)=-\frac{d^2}{dx^2}+q(x)
\end{equation}
with $\mathcal{PT}$-symmetric potential $q(x)$, i.e., $q(x)=\overline{q(-x)}$ were studied by different methods \cite{ABB,CDV,JOU1,HKS,LSZ,MOS2,ZNO1,ZNO3}. In particular, scattering on the $\mathcal{PT}$-symmetric Coulomb potential
was studied on a trajectory of the complex plane \cite{LSZ};  discretization methods were used
for getting the explicit formulae for the reflection and transmission coefficients \cite{ZNO1,ZNO3}; the relationship between $\mathcal{PT}$-symmetric Hamiltonians and reflectionless scattering systems was discussed \cite{ABB, HKS};
spectral singularities were characterized in terms of reflection and transmission coefficients \cite{MOS2}.

If the potential $q(x)$ in (\ref{gere1}) is local, that is, if its support is contained in an interval $(-\rho,\rho)$,
then the corresponding traveling wave functions have the form:
$$
f_1=\left\{\begin{array}{ll}
{e^{-i{k}x}}+R_k^{r}e^{ikx}, & x\geq\rho \\
T_k^{r}e^{-ikx}, & x\leq{-\rho}
\end{array}\right., \ \
f_2=\left\{\begin{array}{ll}
{T_k^l}e^{ikx}, & x\geq\rho \\
e^{i{k}x}+{R_k^l}e^{-ikx}, & x\leq-\rho
\end{array}\right.,
$$
where $R_k^{l}$, $R_k^{r}$ are the left and right reflection coefficients and
$T_k^{l}$, $T_k^{r}$ are the left and right transmission coefficients, respectively and $k>0$.

The $S$-matrix  of (\ref{gere1}) is presented in terms of reflection/transmission coefficients and
the investigation of relationship between the $S$-matrix\footnote{or its counterparts like reflection/transmission coefficients}
and a $\mathcal{PT}$-symmetric operator $H$ generated by (\ref{gere1}) is the subject of the scattering theory.
Examples of such kind of investigations for local $\mathcal{PT}$-symmetric potentials can be found in \cite{CDV,JOU1,MOS2,ZNO1}.

In the present paper we are going to contribute to this inspiring field by studying the scattering of $\mathcal{PT}$-symmetric operators
with the use of an operator-theoretical interpretation of the Lax-Phillips approach in scattering theory \cite{LF} developed in
\cite{KU1,AlAn,Kioto}.

Our choice of the Lax-Phillips approach is explained by the fact that this approach fits the scattering on local potentials well and
it contains an algorithm for the explicit calculation of the analytic continuation of $S$-matrix
in the case of local symmetric potentials.

Another distinctive feature of the Lax-Phillips approach is its
operator-theoretical formulation that allows one to consider the
scattering of many concrete systems with locally supported perturbations from a unique point of view.
Here, the crucial role is played by the concept of $\rho$-perturbed operator (Definition \ref{did2}),
which generalizes the concept of local perturbation to the case of an abstract Hilbert space and
provides simple links to powerful mathematical methods of the extension theory of symmetric operators.
The latter leads to general formula (\ref{red1}) of the analytical continuation of $S$-matrix onto $\mathbb{C}_+$ for
the case of $\rho$-perturbed positive self-adjoint operators (subsection 2.3).
The application of (\ref{red1}) to the case of Schr\"{o}dinger-type differential expression (\ref{gere1})
with symmetric local potential gives the analytical continuation  of
the known $S$-matrix, which is expressed in terms of (generalized) reflection/transmission coefficients
calculated for any $k\in\mathbb{C}_+$ (subsection 2.4).

The results of subsection 2.4 can be considered as a hint about `right definition'  of $S$-matrix for
$\mathcal{PT}$-symmetric $\rho$-perturbed operators. Roughly speaking the idea is to use the general formula
(\ref{red1}) for all $k\in\mathbb{C}_+$, where this formula has sense. We found this definition useful because, for (\ref{gere1}) with
$\mathcal{PT}$-symmetric potentials, it leads to the expressions of $S$-matrices in terms of reflection/transmission coefficients
(subsection 3.3). Moreover, since the $S$-matrix is determined for some subset of $\mathbb{C}_+$ we obtain more informative relationship between reflection/transmission coefficients and $\mathcal{PT}$-symmetric $\rho$-perturbed operators that can be useful for various inverse problems studies.

Let a  $\mathcal{PT}$-symmetric $\rho$-perturbed operator $H$ have the
property of $\mathcal{C}$-symmetry realized by an operator $\mathcal{C}=e^{-Q}\mathcal{P}$ (Definition \ref{neww55})
and let $H$ turn out to be a self-adjoint operator in the Hilbert space $\mathfrak{H}$ endowed with new inner product $(e^{Q}\cdot,\cdot)$.
In that case, we may expect that the corresponding $S$-matrix contains an information about the metric operator $e^Q$.
This gives rise to a problem of recovering of the metric operator $e^Q$ (and, hence the operator $\mathcal{C}$)
by the $S$-matrix of a $\mathcal{PT}$-symmetric $\rho$-perturbed operator.

In this trend, we continue the investigations of \cite{ALBKUZ} and consider the case when the metric operator $e^Q$ fits into the Lax-Phillips scattering structure (subsection 3.4). Theorems \ref{neww78}, \ref{neww144} correspond to the direct problem (metric operator $\to$ the properties of $S$-matrix) and subsection 3.5 contains an example of the successful solution of the inverse problem ($S$-matrix $\to$ recovering the metric operator $e^Q$).

Throughout the paper,  $\mathcal{D}(A)$, $\mathcal{R}(A)$, and $\rho(A)$ denote the domain, the range, and the resolvent set of a linear operator $A$, respectively.
The symbol $A\upharpoonright_{\mathcal{D}}$ means the restriction of $A$ onto a set $\mathcal{D}$.
We denote by ${W^m_2}((a,b),{\mathcal N})$  the Sobolev space of vector functions on $(a,b)$ $(-\infty\leq{a}<b\leq{\infty})$
with values in an auxiliary Hilbert space ${\mathcal N}$; the set $\stackrel{0}{{W}_{2}^{m}}((a,b),{\mathcal N})$ is a subspace of ${W^m_2}((a,b),{\mathcal N})$ determined by the condition: $f\in{\stackrel{0}{{W}_{2}^{m}}((a,b),{\mathcal N})}$ if all derivatives $f^{(k)}(x)$ \ $(k=0,\ldots{m-1})$ vanish at points $x=a, x=b$
(see, e.g., \cite{GG} for detail).

\section{Elements of Lax--Phillips Scattering Theory}

\subsection{Definition of unperturbed and $\rho$-perturbed operators.}
 Let $B$ be a densely defined symmetric operator in a Hilbert space $\mathfrak{H}$ with inner product $(\cdot,\cdot)$.
The defect numbers $m_\pm$ of $B$ are defined as $m_\pm=\dim\ker(B^*\pm{i}I)$, where
$B^*$ is the adjoint operator of $B$.

 A symmetric operator $B$ is called \emph{simple}  if it does not induce a self-adjoint
operator in any proper subspace of ${\mathfrak H}$ and $B$ is called \emph{maximal symmetric} if one of its defect numbers $m_{\pm}$
is equal to zero. In what follows, we suppose that $m_-=0$, i.e.,  $m=m_+=\dim\ker(B^*+{i}I)>0$.

It is known \cite{AkG1} that a simple maximal symmetric operator $B$ with nonzero defect number $m$
in $\mathfrak{H}$ \emph{is unitarily equivalent} to a simple maximal symmetric operator
\begin{equation}\label{neww45}
\mathcal{B}=i\frac{d}{dx}, \qquad \mathcal{D}(\mathcal{B})=\{u\in{W^1_2}({\mathbb R}_+,{\mathcal N}) : u(0)=0\}
\end{equation}
in the Hilbert space $L_2({\mathbb R}_+,{\mathcal N})$, where $\mathbb{R}_+=\{x\in\mathbb{R}\ : \ x\geq0\}$ and the dimension of the auxiliary Hilbert space ${\mathcal N}$ coincides with $m$.  This relationship immediately leads to the following statement.
  \begin{lemma}\label{l1}
 The operator $B^2$ is a closed densely defined symmetric operator in $\mathfrak{H}$ and
 ${B^2}^{*}={B^{*}}^2.$
\end{lemma}
\begin{definition}\label{did1}
A self-adjoint extension $H$ of $B^{2}$ is called \emph{unperturbed} if
    \begin{equation}\label{e5}
    (Hf,f)={\Vert B^{*}f \Vert}^{2}, \qquad {f}\in{\mathcal{D}(H)}.
    \end{equation}
  \end{definition}

 Every unperturbed operator $H$ has a purely absolutely continuous spectrum with the same multiplicity $m$ at
 each point of $[0,\infty)$, where $m$ is the non-zero defect number of $B$ \cite{KU1}.

Since $B$ is unitarily equivalent to the operator $\mathcal{B}$ defined by (\ref{neww45}),
the semigroup $V(t)=e^{iBt}$ is a completely nonunitary semigroup of
isometries \cite[Theorem 9.3, Chapter III]{SNK}

Denote $\mathfrak{H}_{\rho}=V(\rho)\mathfrak{H}$ \ $(\rho\geq{0})$.
We have $\mathfrak{H}_{\rho_1}\supset\mathfrak{H}_{\rho_2}$ for $0\leq\rho_1<\rho_2$.

The restriction of $B$ on $V(\rho)\mathcal{D}(B)$ gives rise to the
simple maximal symmetric operator
\begin{equation}\label{nee3}
B_{\rho}:=B\upharpoonright_{\mathcal{D}(B_{\rho})}, \qquad    \mathcal{D}(B_{\rho})=V(\rho)\mathcal{D}(B)
\end{equation}
acting in the Hilbert space $\mathfrak{H}_{\rho}$.
By Lemma \ref{l1},  $B_\rho^2$ is a densely defined symmetric operator in $\mathfrak{H}_{\rho}$.

\begin{definition}\label{did2}
 A closed densely defined operator $H$ in $\mathfrak{H}$ is called  \emph{$\rho$-perturbed} if
 $H$ and its adjoint operator $H^*$ are extensions of $B_{\rho}^2$, i.e.,
 $$
 Hu={B_{\rho}^2}u \quad \mbox{and} \quad H^*u={B_{\rho}^2}u \qquad \mbox{for all} \quad  u\in\mathcal{D}(B_{\rho}^2).
 $$
\end{definition}

{\bf Example 1.}
 Let $\mathfrak{H}=L_2(\mathbb{R})$ and let $\mathbb{R}_\pm=\{x\in\mathbb{R}\ : \ {\pm}x>0\}$. The operator
$$
B=(\textsf{sgn}\ x)i\frac{d}{dx}, \qquad \mathcal{D}(B)=\stackrel{0}{{W}_{2}^{1}}(\Bbb R_{-})\oplus\stackrel{0}{{W}_{2}^{1}}(\Bbb R_{+})
$$
is a simple maximal symmetric operator in $L_2(\mathbb{R})$ and
$$
B^2=-\frac{d^2}{dx^2}, \qquad \mathcal{D}(B^2)=\stackrel{0}{{W}_{2}^{2}}(\Bbb R_{-})\oplus\stackrel{0}{{W}_{2}^{2}}(\Bbb R_{+}).
$$
The operator
$$
H=-\frac{d^2}{dx^2}, \qquad \mathcal{D}(H)=W^2_2(\mathbb{R})
$$
is a self-adjoint extension of $B^2$ and it satisfies (\ref{e5}). Therefore, $H$ is an unperturbed operator.

Another unperturbed operator is the Friedrichs extension $H_\mu=B^*B$ of $B^2$:
\begin{equation}\label{ses121}
 H_\mu=-\frac{d^2}{dx^2}, \quad \mathcal{D}(H_\mu)=\{u\in{W}^2_2(\mathbb{R}_-)\oplus{W}^2_2(\mathbb{R}_+)  :  u(0-)=u(0+)=0\}
\end{equation}

The isometric semigroup $V(t)=e^{iBt}$ ($t\geq0$) acts as follows
$$
V(t)f(x)=\left\{\begin{array}{ll}
f(x-t), & \quad x\geq{t}, \\
0, & \quad |x|<t, \\
f(x+t), & \quad x\leq{-t}
\end{array}\right.
$$

This means that  $\mathfrak{H}_\rho=L_2(\mathbb{R}\setminus(-\rho,\rho))$,
\begin{equation}\label{ses124}
B_\rho=(\textsf{sgn}\ x)i\frac{d}{dx}, \qquad \mathcal{D}(B_\rho)=\stackrel{0}{{W}_{2}^{1}}(-\infty,-\rho)\oplus\stackrel{0}{{W}_{2}^{1}}(\rho,\infty),
\end{equation}
and
\begin{equation}\label{ses1}
B_\rho^2=-\frac{d^2}{dx^2}, \qquad \mathcal{D}(B^2_\rho)=\stackrel{0}{{W}_{2}^{2}}(-\infty,-\rho)\oplus\stackrel{0}{{W}_{2}^{2}}(\rho, \infty). \end{equation}

Consider the differential expression
\begin{equation}\label{rada1}
l(\cdot)=-\frac{d^2}{dx^2}+q(x), \qquad \textsf{supp}\ q(x)\subset(-\rho,\rho), \quad x\in\mathbb{R}
\end{equation}
and suppose that $l(\cdot)$ determines a closed densely defined operator $H$ in $L_2(\mathbb{R})$. Then, the operator
$H$ and its adjoint operator $H^*$ are extensions of the symmetric operator
$B_\rho^2$ defined by (\ref{ses1}). Therefore, due to Definition \ref{did2}, the operator $H$ is $\rho$-perturbed.

\begin{remark}
Usually \cite{LF}, the Lax-Phillips perturbed and unperturbed evolutions are defined with the use of incoming $D_-$ and outgoing $D_+$ subspaces  for the unitary group of solutions $W_H(t)$ of the Cauchy problem for the operator differential equation
\begin{equation}\label{bonn60}
\frac{d^2}{dt^2}u=-Hu,
\end{equation}
where $H$ is a positive (i.e. $(Hf,f)>0$ for all nonzero $f\in\mathcal{D}(H)$) self-adjoint operator in $\mathfrak{H}$.
In particular, the existence of orthogonal subspaces $D_\pm$ of the space of Cauchy data\footnote{the Hilbert space $\mathfrak{H}_H$ is
a completion of $\mathcal{D}(H)$ with respect to the norm $\|\cdot\|_H=(H\cdot,\cdot)$}  $\mathfrak{G}=\mathfrak{H}_H\oplus\mathfrak{H}$
with the properties
$$
\begin{array}{l}
 (i) \quad  W_H(-t)D_{-}\subset{D_{-}}, \qquad  W_H(t)D_{+}\subset{D_{+}}, \qquad t\geq0; \vspace{3mm} \\
 (ii) \quad \bigcap_{t\geq 0}W_H(-t)D_{-}=\bigcap_{t\geq 0}W_H(t)D_{+}=\{0\},
 \end{array}
$$
characterizes the perturbed evolution. The unperturbed evolution is determined by the additional requirement
$$
(iii) \quad D_-\oplus{D_+}=\mathfrak{G}.
$$

These definitions are coordinated with Definitions \ref{did1}, \ref{did2} in the following sense:
if a positive self-adjoint operator $H$ on the right-hand side of (\ref{bonn60}) is $\rho$-perturbed or unperturbed, then
the corresponding group $W_H(t)$ of Cauchy problem solutions possesses
orthogonal subspaces $D_\pm$ with the properties (i)-(ii) or (i)-(iii), respectively \cite{AlAn}.
\end{remark}

\subsection{Definition and properties of $S$-matrix.}
It is easy to check that the relation (\ref{e5}) holds for the Friedrichs extension $H_\mu=B^*B$ of $B^2$.
Therefore, the Friedrichs extension $H_\mu$ remains an unperturbed operator for any choice of $B$.

\begin{proposition}[\cite{Kioto}]\label{pepe1}
Let $H$ be a positive self-adjoint $\rho$-perturbed operator for a given simple maximal symmetric
operator $B$ in $\mathfrak{H}$. Then the wave operators
$$
\Omega_{\pm}(H, H_\mu):=s-\lim_{t\to\pm\infty}e^{iHt}e^{-iH_{\mu}t}
$$
exist and are isometric in ${\mathfrak H}$.
\end{proposition}

The operator $S_{(H,H_\mu)}=\Omega_+^*(H, H_\mu)\Omega_-(H, H_\mu)$
is called \emph{the scattering operator}.

The properties of $S_{(H,H_\mu)}$ is feeling better in terms of the spectral representation of an unperturbed operator \cite{BM1}.
To this end we construct the spectral representation of $H_\mu$.

Since $B$ is unitarily equivalent to the operator $\mathcal{B}=i\frac{d}{dx}$ defined by (\ref{neww45}),
there exists a unitary mapping $Y : \mathfrak{H}\to{L_2}({\mathbb R}_+,{\mathcal N})$ such that
\begin{equation}\label{p21}
B=Y^{-1}{\mathcal B}Y, \quad \mathcal{D}(B)=Y^{-1}\mathcal{D}(\mathcal{B})=Y^{-1}\{u\in{W^1_2}({\mathbb R}_+,{\mathcal N})  :  u(0)=0\}.
\end{equation}
Then the operator
$$
({\mathcal F}f)(\delta)=\sqrt{\frac{2}{\pi}}\int_0^\infty{\sin}{\delta{x}}(Yf)(x)dx, \qquad
{f}\in\mathfrak{H}, \ \ \delta>0
$$
isometrically maps $\mathfrak{H}$ onto ${L}_2({\mathbb R}_+,{\mathcal N})$ and
$$
({\mathcal F}B^*Bf)(\delta)=\delta^2({\mathcal F}f)(\delta), \qquad {f}\in\mathcal{D}(B^*B).
$$
The mapping $\mathcal{F}$ determines a
spectral representation $L_2({\mathbb R}_+,{\mathcal N})$ of $H_\mu=B^*B$ in which the action of $H_\mu$ corresponds to
the multiplication by the modified spectral parameter $\delta^2$.

 The image ${\mathbb S}={\mathcal F}S_{(H,H_\mu)}{\mathcal F}^{-1}$ of the scattering operator
 $S_{(H,H_\mu)}$ in the spectral representation ${L}_2({\mathbb R}_+,{\mathcal N})$
 can be realized as the multiplication by an operator-valued
 function ${\mathbb S}(\delta)$, the values of which
 are bounded operators in ${\mathcal N}$ for almost all $\delta\in\mathbb{R}_+$.
Precisely,
$$
{\mathbb S}f={\mathbb S}(\delta)f(\delta), \qquad {f}\in{L_2({\mathbb R}_+,{\mathcal N})}.
$$

Let us extend the function ${\mathbb S}(\delta)$ onto the whole real axis
$$
{\mathbb S}(-\delta):={\mathbb S}^*(\delta), \qquad \delta>0,
$$
where ${\mathbb S}^*$ means the adjoint operator in ${\mathcal N}$.

The obtained operator-valued function ${\mathbb S}(\cdot)$ depends on the choice of an
auxiliary space ${\mathcal N}$ in (\ref{p21}). However, for any choice of ${\mathcal N}$,
the function ${\mathbb S}(\cdot)$ is the boundary value\footnote{in the sense of strong convergence in ${\mathcal N}$} of an analytic
function ${\mathbb S}(k)$ in $\mathbb{C}_+=\{k\in\mathbb{C} : \textsf{Im} \ k>0\}$.
The values of ${\mathbb S}(k)$  are contraction operators in ${\mathcal N}$
(see  \cite[Theorems 4.1, 4.2]{AlAn}, \cite[Theorem 2.3]{Kioto} for details).

The operator-valued function ${\mathbb S}(\cdot)$ is called the $S$-{\it matrix} of
the positive self-adjoint $\rho$-perturbed operator  $H$.

\subsection{An operator method for the calculation of the $S$-matrix.}
The $S$-matrix ${\mathbb S}(\cdot)$ contains information about the $\rho$-perturbed operator ${H}$
and the investigation of the relationship between
$\mathbb{S}(\cdot)$ and ${H}$ is the proper subject of Lax-Phillips scattering theory.

The common feature of  unperturbed $H_\mu$ and $\rho$-perturbed ${H}$ operators is that
 \emph{they are extensions of
a given symmetric operator} $B^2_\rho$. This leads to a simple recipe for finding $\mathbb{S}(\cdot)$ \cite{AlAn,Kioto}.
We begin with the following auxiliary result.
\begin{lemma}\label{meme2}
Let a closed densely defined operator $H$ be $\rho$-perturbed in the sense of Definition \ref{did2}. Then
\begin{equation}\label{aaa1}
{P_\rho}\mathcal{D}({H})\subset\mathcal{D}({B^*_\rho}^2) \quad \mbox{and} \quad P_\rho{H}{f}={B^*_\rho}^2P_\rho{f} \quad \forall{f}\in\mathcal{D}({H}),
\end{equation}
where $P_\rho$ is the orthogonal projection of ${\mathfrak{H}}$ onto $\mathfrak{H}_\rho$.
\end{lemma}
\emph{Proof.} If $H$ is $\rho$-perturbed, then $H\supset{B^2_\rho}$ and $H^*\supset{B^2_\rho}$. Hence,
\begin{eqnarray*}
(P_\rho{H}{f},u)=({H}{f},u)=(f,H^*u)=({f},B^2_{\rho}u)= & &  \\
 (P_\rho{f},B^2_\rho{u})=({B^2_\rho}^*P_\rho{f},{u})=({{B^*_\rho}^2}P_\rho{f},{u}) & &
\end{eqnarray*}
for all ${f}\in\mathcal{D}({H})$ and for all ${u}\in\mathcal{D}(B^2_\rho)$. The obtained relation
implies (\ref{aaa1}). \rule{2mm}{2mm}

Lemma \ref{meme2} shows that the operators
\begin{equation}\label{ada52}
H_k={B^*_\rho}^2\upharpoonright_{\mathcal{D}(H_k)}, \qquad \mathcal{D}(H_k)=P_\rho({H}-k^2{I})^{-1}\mathfrak{H}_\rho
\end{equation}
are well defined in $\mathfrak{H}_\rho$ for all $k\in\Lambda_+=\{k\in\mathbb{C}_+ : k^2\in\rho(H)\}$.
\begin{definition}\label{did56}
The set of operators $\{H_k\}_{k\in\Lambda_+}$ acting in $\mathfrak{H}_\rho$ and determined by (\ref{ada52}) is called \emph{the image set} of
a $\rho$-perturbed operator $H$.
\end{definition}

If a $\rho$-perturbed operator $H$ is a positive self-adjoint operator in $\mathfrak{H}$, then $\Lambda_+=\mathbb{C}_+$ and
operators $H_k$ from the image set are defined for all $k\in\mathbb{C}_+$.

It is useful to describe operators $H_k$ in terms of a boundary triplet
$({\mathcal H},\Gamma_{0},\Gamma_{1})$ of $B^2_\rho$ defined as follows.

Let ${\mathcal H}=\kerr({B^*_\rho}^2+I)$. Then $\mathcal{D}({B^*_\rho}^2)=\mathcal{D}(B_\rho^*B_\rho)\dot{+}{\mathcal H}$ and hence,
every function $f\in\mathcal{D}({B^*_\rho}^2)$ is uniquely decomposed:
\begin{equation}\label{bonn41}
f=u+h, \qquad u\in{D({B^*_\rho}{B_\rho})}, \quad h\in{\mathcal H}.
\end{equation}

The decomposition (\ref{bonn41}) allows to define the linear mappings $\Gamma_{0}$ and $\Gamma_{1}$ from
$\mathcal{D}({B^*_\rho}^2)$ into ${\mathcal H}$:
 \begin{equation}\label{e7}
 \Gamma_{0}f=\Gamma_{0}(u+h)=h,  \qquad \Gamma_{1}f=\Gamma_{1}(u+h)=P_{\mathcal H}(B_\rho^*B_\rho+I)u,
    \end{equation}
  where $P_{\mathcal H}$ is the orthogonal projector of ${\mathfrak H}_\rho$ onto the subspace ${\mathcal H}$.

The triple $({\mathcal H},\Gamma_{0},\Gamma_{1})$ is called the positive boundary triplet (positive boundary value space) of $B_\rho^2$ \cite{GG}.
\begin{lemma}[\cite{AlAn}]\label{pepe15}
Let $H$ be a positive self-adjoint  $\rho$-perturbed operator. Then
the operators $H_k$ from the image set $\{H_k\}_{k\in\mathbb{C}_+}$
are restrictions of ${B^*_\rho}^2$ onto
\begin{equation}\label{sese2}
  \mathcal{D}(H_k)=\{f\in{\mathcal{D}({B^*_\rho}^2)} \ :  \  T_k\Gamma_{1}f=\Gamma_{0}f\},
 \end{equation}
where $T_k$ are bounded operators in the Hilbert space ${\mathcal H}=\ker({B^*_\rho}^2+I)$ and
$T_k^*=T_{-\overline{k}}$. Furthermore, the operator $T_k$ is maximal dissipative (accumulative) when $\textsf{Re} \ k>0$ ($\textsf{Re} \ k<0$) and $T_k$ is a nonnegative self-adjoint operator with $\|T_k\|\leq{1/2}$ while $\textsf{Re} \ k=0$.
\end{lemma}

 It follows from (\ref{p21}) that the dimension of ${\mathcal H}$ coincides with the dimension of
 the auxiliary space ${\mathcal N}$ in the definition of the $S$-matrix ${\mathbb S}(\cdot)$.
 Let us identify ${\mathcal N}$ with ${\mathcal H}$ for the simplicity.

\begin{theorem}[\cite{AlAn}]\label{esse3}
Let $H$ be a positive self-adjoint  $\rho$-perturbed operator. Then the
$S$-matrix ${\mathbb S}(\cdot)$ is an analytic operator-valued function
in $\mathbb{C}_+$ and
\begin{equation}\label{red1}
{\mathbb S}(k)=[I-2(1-ik)T_k][I-2(1+ik)T_k]^{-1}, \quad k\in\mathbb{C}_+.
\end{equation}
where $T_k$ are taken from (\ref{sese2}).
\end{theorem}
\begin{remark}\label{esse3d}
The properties of $T_k$ described in Lemma \ref{pepe15}
yield that the formula (\ref{red1}) determines an analytic contraction-valued function in $\mathbb{C}_+$, which satisfies the relation ${\mathbb S}(-\overline{k})={\mathbb S}^*(k)$, where ${\mathbb S}^*(k)$ is the adjoint operator of  ${\mathbb S}(k)$ with respect to the inner product
$(\cdot,\cdot)$ in $\mathcal{H}$.
\end{remark}

In Theorem \ref{esse3}, the auxiliary space ${\mathcal N}$ in the spectral representation $L_2(\mathbb{R}, {\mathcal N})$
is identified with $\mathcal{H}$. It looks natural to rewrite the obtained result for the general case.

Let $Y$ be an isometric mapping of $\mathfrak{H}$ onto $L_2(\mathbb{R}_+,{\mathcal N})$ from (\ref{p21}).
Taking (\ref{nee3}) into account, we conclude that $Y$ maps $\mathcal{H}=\kerr({B_\rho^*}^2+I)$ onto the subspace
$\{e^{-x}u : {u}\in{{\mathcal N}}\}$ of ${L_2((\rho,+\infty),{\mathcal N})}$.
Identifying functions $f(x)={e^{-x}u}$ with elements $v={\alpha}u\in{{\mathcal N}}$:
\begin{equation}\label{rest1}
f(x)=e^{-x}u \leftrightarrow v \ (={\alpha}u), \qquad \alpha=\frac{1}{\sqrt{2}}e^{-\rho}
\end{equation}
where the constant $\alpha$ is chosen in such a way that
$\|f\|_{L_2}=\|v\|_{{\mathcal N}}$,
we obtain that $Y$ isometrically maps $\mathcal{H}$ onto ${\mathcal N}$.

For the operator-valued function
${\textsf S}(\cdot)=Y{\mathbb S}(\cdot)Y^{-1}$,
Theorem \ref{esse3} is reformulated as follows:
\begin{theorem}\label{esse3b}
The $S$-matrix ${\textsf S}(\cdot)$ of $H$ has the form
\begin{equation}\label{red1b}
{\textsf S}(k)=[I-2(1-ik){\textsf T}_k][I-2(1+ik){\textsf T}_k]^{-1}, \  k\in\mathbb{C}_+,
\end{equation}
where ${\textsf T}_k$ characterizes the domain $\mathcal{D}(H_k)$:
\begin{equation}\label{sese2b}
  \mathcal{D}(H_k)=\{f\in{\mathcal{D}({B^*_\rho}^2)} \ :  \  {\textsf T}_kY\Gamma_{1}f=Y\Gamma_{0}f\}.
 \end{equation}
\end{theorem}

\subsection{Application to Schr\"{o}dinger operator with local potential.}
Let the differential expression (\ref{rada1})
determine a positive self-adjoint operator $H$ in $\mathfrak{H}=L_2(\mathbb{R})$.
Then $H$ is $\rho$-perturbed in the sense of Definition \ref{did2}, where $B_\rho^2$ is defined by (\ref{ses1})
(see Example 1).
Hence, we can find the $S$-matrix ${\textsf S}(\cdot)$ of $H$ in $\mathbb{C}_+$.

A simple calculation with the use of (\ref{ses121}) and (\ref{ses1})  leads to the conclusion that the
positive boundary triplet $({\mathcal H},\Gamma_{0},\Gamma_{1})$ of $B_\rho^2$  defined by formulas (\ref{bonn41}) and (\ref{e7})
has the following form: the space $\mathcal{H}$ coincides with the linear span of
functions
$$
\psi_-(x)=\left\{\begin{array}{ll}
0,  & x\geq\rho \\
2e^{\rho+x}, &  x\leq{-\rho}
\end{array}\right. \quad \mbox{and} \quad
\psi_+(x)=\left\{\begin{array}{ll}
2e^{\rho-x}, &  x\geq\rho \\
0, &  x\leq{-\rho}
\end{array}\right. ;
$$
the operators $\Gamma_0: \mathcal{D}({B^*_\rho}^2)\to\mathcal{H}$ act as follows
\begin{equation}\label{meme1}
\begin{array}{l}
\Gamma_0f=\frac{1}{2}(f(-\rho)\psi_-+f(\rho)\psi_+), \vspace{3mm} \\
\Gamma_1f=[f(-\rho)-f'(-\rho)]\psi_-+[f(\rho)+f'(\rho)]\psi_+,
\end{array}
\end{equation}
where $f\in\mathcal{D}({B^*_\rho}^2)={{W}_{2}^{2}}(-\infty,-\rho)\oplus{{W}_{2}^{2}}(\rho, \infty)$.

Let us choose $\mathbb{C}^2$ as the auxiliary space ${\mathcal N}$ and denote elements of its canonical basis by  $e_+=(1,0)^T$ and $e_-=(0,1)^T$.

The operator $Y$ defined by the formulas:
\begin{equation}\label{deder8}
\begin{array}{ll}
Yf(x)=f(x){e_+}, \ x\in\mathbb{R}_+  & \mbox{for all} \ {f}\in{L_2(\mathbb{R}_+)} \vspace{2mm} \\
Yg(x)=g(-x){e_-}, \ x\in\mathbb{R}_- & \mbox{for all} \ {g}\in{L_2(\mathbb{R}_-)}
\end{array}
\end{equation}
maps isometrically $L_2(\mathbb{R})$ onto $L_2(\mathbb{R}_+, \mathbb{C}^2)$ and it satisfies relation (\ref{p21}).
Moreover, taking (\ref{rest1}) into account, we get $Y\psi_\pm=\sqrt{2}e_{\pm}$. The obtained relation and the explicit formulas (\ref{meme1}) for $\Gamma_j$ allow us to reinterpret (\ref{sese2b}) as follows: the domain $\mathcal{D}(H_k)$ consists of those functions  $f\in{{W}_{2}^{2}}(-\infty,-\rho)\oplus{{W}_{2}^{2}}(\rho, \infty)$ for which
\begin{equation}\label{sese2c}
{\textsf T}_k\left(\begin{array}{c}
f(\rho)+f'(\rho) \\
f(-\rho)-f'(-\rho)
\end{array}\right)=\frac{1}{2}\left(\begin{array}{c}
f(\rho) \\
f(-\rho)
\end{array}\right).
\end{equation}

The matrix ${\textsf T}_k=\|t_{ij}\|_{ij}^2$ in (\ref{sese2c}) is completely determined
by two linearly independent functions $f_1, f_2\in\mathcal{D}(H_k)\setminus\mathcal{D}(B^2_\rho)$.
Precisely, we have to know values $f_j(\pm\rho)$ together with values of derivatives $f_j'(\pm\rho)$.

Since $\mathfrak{H}_\rho=\mathcal{R}(B^2_\rho-k^2I)\oplus\ker({B^*_\rho}^2-{\overline{k}}^2I)$, the second relation in
(\ref{ada52}) leads to the conclusion that
\begin{equation}\label{rest4}
\mathcal{D}(H_k)=\mathcal{D}(B^2_\rho)\dot{+}P_\rho(H-k^2I)^{-1}\ker({B^*_\rho}^2-{\overline{k}}^2I),
\end{equation}
where $P_\rho$ is the orthogonal projection onto $\mathfrak{H}_\rho=L_2(\mathbb{R}\setminus(-\rho,\rho))$.

Let us assume that $k\in\mathbb{C}_+'=\mathbb{C}_+\setminus{i\mathbb{R}_+}=\{k\in\mathbb{C}_+ : \textsf{Re} \ k\not=0\}$. Then the functions
$$
h_1=\left\{\begin{array}{l}
\beta{e^{-i\overline{k}x}}, \quad x\geq\rho \\
0, \quad x<\rho
\end{array}\right., \quad
h_2=\left\{\begin{array}{l}
0, \quad x>-\rho \\
\beta{e^{i\overline{k}x}}, \quad x\leq-\rho
\end{array}\right., \qquad \beta={\overline{k}}^2-k^2
$$
form a basis of $\ker({B^*_\rho}^2-{\overline{k}}^2I)$.
(The coefficient $\beta$ is used for the simplification of formulas below.)

Let $\tilde{f}_j\in\mathcal{D}(H)$ be solutions of equations
\begin{equation}\label{rest1963}
(H-k^2I)\tilde{f}_j=h_j, \qquad j=1,2.
\end{equation}
It follows from  (\ref{rest4}) and (\ref{rest1963}) that the functions $f_j=P_\rho{\tilde{f}_j}$ belong to $\mathcal{D}(H_k)\setminus\mathcal{D}(B^2_\rho)$
and they are linearly independent. Furthermore, taking (\ref{rada1}) into account, we get
\begin{equation}\label{deder1}
f_1=\left\{\begin{array}{ll}
{e^{-i\overline{k}x}}+R_k^{r}e^{ikx}, & x\geq\rho \\
T_k^{r}e^{-ikx}, & x\leq{-\rho}
\end{array}\right., \ \
f_2=\left\{\begin{array}{ll}
{T_k^l}e^{ikx}, & x\geq\rho \\
e^{i\overline{k}x}+{R_k^l}e^{-ikx}, & x\leq-\rho
\end{array}\right.,
\end{equation}
where  $R_k^{l}$, $R_k^{r}$ are the left and right reflection coefficients and
$T_k^{l}$, $T_k^{r}$ are the left and right transmission coefficients, respectively.

The formulas (\ref{deder1}) allow us to express the values of $f_j(\pm\rho)$ and $f'_j(\pm\rho)$
in terms of coefficients $R_k^{l}$, $R_k^{r}$, $T_k^{l}$ and $T_k^{r}$.
Substituting these values into (\ref{sese2c}) and making elementary calculations we determine the entries
of ${\textsf T}_k=\|t_{ij}\|_{ij}^2$. Then, the formula (\ref{red1b}) leads to the following expression
of the $S$-matrix ${\textsf S}(\cdot)$:
\begin{equation}\label{rest2}
{\textsf S}(k)=-e^{2i\rho\textsf{Re} k}\frac{k}{\textsf{Re}\ k}\left(\begin{array}{cc}
R_k^r+\displaystyle{e^{-2i\rho\textsf{Re} k}\frac{\textsf{Im}\ k}{k}} & T_k^l  \vspace{4mm} \\
T_k^r & R_k^l+\displaystyle{e^{-2i\rho\textsf{Re} k}\frac{\textsf{Im}\ k}{k}}
\end{array}
\right)
\end{equation}
The formula (\ref{rest2}) is obtained for $k\in\mathbb{C}_+'$ and it can be extended onto $\mathbb{C}_+$ by the continuity.
For real $k$ the expression (\ref{rest2}) is reduced to
\begin{equation}\label{rest2b}
{\textsf S}(k)=-e^{2i\rho{k}}\left(\begin{array}{cc}
R_k^r & T_k^l  \vspace{4mm} \\
T_k^r & R_k^l
\end{array}
\right).
\end{equation}

\section{Scattering for $\mathcal{PT}$-symmetric operators}
\subsection{Definition of $\mathcal{PT}$-symmetric operators}
Let $\mathfrak{H}$ be a Hilbert space with inner product
$(\cdot,\cdot)$. A linear operator $\mathcal{P}$ in $\mathfrak{H}$ is called \emph{unitary involution} if
\begin{equation}\label{bonn1}
(i) \quad \mathcal{P}^2=I, \qquad (ii) \quad (\mathcal{P}f,\mathcal{P}g)=(f,g), \quad {f,g}\in\mathfrak{H}.
\end{equation}

A modification of condition (ii) in (\ref{bonn1}) leads to the definition of the conjugation operator.
An operator $\mathcal{T}$ in $\mathfrak{H}$ is called \emph{conjugation} if
\begin{equation}\label{bonn2}
(i) \quad \mathcal{T}^2=I, \qquad (ii) \quad (\mathcal{T}f,\mathcal{T}g)=(g,f), \quad {f,g}\in\mathfrak{H}.
\end{equation}
The conjugation operator is bounded in $\mathfrak{H}$ but, in contrast to the case of an involution, $\mathcal{T}$ is \emph{anti-linear} in the sense that
\begin{equation}\label{bonn2b}
 \mathcal{T}(\alpha{f}+\beta{g})=\overline{\alpha}\mathcal{T}{f}+\overline{\beta}\mathcal{T}{g}, \qquad \alpha, \beta\in\mathbb{C}, \quad f,g\in\mathfrak{H}.
\end{equation}

Let us fix a unitary involution $\mathcal{P}$ and a conjugation $\mathcal{T}$ in $\mathfrak{H}$ assuming in what follows that
$\mathcal{P}$ and $\mathcal{T}$ \emph{commute}, that is, $\mathcal{P}\mathcal{T}=\mathcal{T}\mathcal{P}$.
This means that $\mathcal{P}\mathcal{T}$ is also a conjugation.

\begin{definition}\label{ddd1}
A closed densely defined linear operator $H$ in $\mathfrak{H}$ is called $\mathcal{P}\mathcal{T}$-symmetric if the relation
\begin{equation}\label{bonn3}
\mathcal{P}\mathcal{T}Hf=H\mathcal{P}\mathcal{T}f
\end{equation}
holds for all elements $f\in\mathcal{D}(H)$.
\end{definition}
\begin{remark}
In what follows we will often use operator identities like
\begin{equation}\label{neww11}
XA=BX,
\end{equation}
where $A$ and $B$ are (possible) unbounded operators in a Hilbert space $\mathfrak{H}$ and  $X$ is a bounded operator in $\mathfrak{H}$.
In that case, we \emph{always assume that (\ref{neww11}) holds on $\mathcal{D}(A)$}. This means that
$X\mathcal{D}(A)\subset\mathcal{D}(B)$ and the identity $XAu=BXu$ holds for all $u\in\mathcal{D}(A)$.
If $A$ is bounded, then (\ref{neww11}) should hold on the whole $\mathfrak{H}$.
\end{remark}

\begin{lemma}[\cite{ALBKUZ}]\label{abbb1}
If $H$ is $\mathcal{P}\mathcal{T}$-symmetric in a Hilbert space $\mathfrak{H}$, then its adjoint operator $H^*$ is also $\mathcal{P}\mathcal{T}$-symmetric.
\end{lemma}

\subsection{The $S$-matrix for $\mathcal{PT}$-symmetric $\rho$-perturbed operators.}
We begin with the following auxiliary result.
\begin{lemma}\label{neww14}
Let $B$ commute with $\mathcal{P}$
and anti-commute with $\mathcal{T}$
\begin{equation}\label{bebe3}
\mathcal{P}B=B\mathcal{P}, \qquad  \mathcal{T}B=-B\mathcal{T},
\end{equation}
and let $H$ be a $\mathcal{PT}$-symmetric $\rho$-perturbed operator.
Then the operators $H_k$ from the image set $\{H_k\}_{k\in\Lambda_+}$ of $H$ (see Definition \ref{did56}) satisfy the relation
$$
\mathcal{PT}H_k=H_{-\overline{k}}\mathcal{PT}.
$$
\end{lemma}
\emph{Proof.} It follows from (\ref{bonn2b}) and (\ref{bebe3}) that $V(t)=e^{iBt}$ commutes with $\mathcal{PT}$. Hence, the subspace $\mathfrak{H}_\rho=V(\rho)\mathfrak{H}$ reduces $\mathcal{PT}$ and the orthogonal projection $P_\rho$
onto $\mathfrak{H}_\rho$ commutes with $\mathcal{PT}$. Furthermore
\begin{equation}\label{red14}
B_{\rho}{\mathcal{PT}}=-{\mathcal{PT}}B_\rho, \qquad B_\rho^*{\mathcal{PT}}=-{\mathcal{PT}}B_\rho^*,
\end{equation}
due to (\ref{nee3}) and (\ref{bebe3}).
Hence, $\mathcal{PT}{B^*_\rho}^2={B^*_\rho}^2\mathcal{PT}$ and
$$
\mathcal{PT}\mathcal{D}(H_k)=\mathcal{PT}P_\rho({H}-k^2{I})^{-1}\mathfrak{H}_\rho=P_\rho({H}-{\overline{k}}^2{I})^{-1}\mathcal{PT}\mathfrak{H}_\rho=\mathcal{D}(H_{-\overline{k}}).
$$
Combining the obtained relations with (\ref{ada52}) we complete the proof.
\rule{2mm}{2mm}

\smallskip

Let us suppose that the operators $H_k$ from the image set $\{H_k\}_{k\in\Lambda_+}$ of a $\mathcal{PT}$-symmetric
$\rho$-perturbed  operator $H$ \emph{can be determined by the relation (\ref{sese2}) with
bounded operators $T_k$ in $\mathcal{H}$}. Then we can formally define
\emph{the $S$-matrix of $H$} by the formula
(\ref{red1}) for all $k\in{\Lambda_+}$ such that $0\in\rho(I-2(1+ik)T_k)$.

\smallskip

\emph{We will consider the operator-valued function ${\mathbb S}(\cdot)$ defined in such a way (as well as its image
${\textsf{S}}(\cdot)=Y{\mathbb S}(k)Y^{-1}$ in ${\mathcal N}$) as
the $S$-matrix of a $\mathcal{PT}$-symmetric $\rho$-perturbed operator $H$.}

\smallskip

Of course, in contrast to the case of positive self-adjoint $\rho$-perturbed operators considered in Section 2,
our definition is rather formal. In particular, we do not take care about the existence of wave operators and other important auxiliary things.
However, we found this definition useful because it provides an explicit relation to the image set $\{H_k\}$ of $H$ that may be important for
the inverse problem studies.

\begin{proposition}\label{neww28}
Let $B$ satisfy (\ref{bebe3}) and let \ ${\mathbb S}(\cdot)$ be the $S$-matrix of $\mathcal{PT}$-symmetric
$\rho$-perturbed operator $H$. Then
$$
\mathcal{PT}{\mathbb S}(k)={\mathbb S}(-\overline{k})\mathcal{PT}
$$
for those $k\in\Lambda_+$ that \ ${\mathbb S}(k)$ exists.
\end{proposition}
\emph{Proof.} First of all, we show that the existence of ${\mathbb S}(k)$ implies
the existence of ${\mathbb S}(-\overline{k})$.
Indeed, the existence of ${\mathbb S}(k)$ is equivalent to the following conditions:
\begin{enumerate}
  \item[(i)] \quad the corresponding operator $T_k$  in (\ref{sese2}) is bounded; \vspace{1mm}
  \item[(ii)]  \quad  $0\in\rho(I-2(1+ik)T_k)$.
\end{enumerate}

 It follows from (i) and \cite[Theorem 2.2, Chapter 3]{GG} that $-1\in\rho(H_k)$ and
\begin{equation}\label{bebe45}
T_k=(H_k+I)^{-1}-(B^*_\rho{B_\rho}+I)^{-1}.
\end{equation}
 The relations (\ref{red14}) imply that $\mathcal{PT}B^*_\rho{B_\rho}=B^*_\rho{B_\rho}\mathcal{PT}$. Hence,
$$
\mathcal{PT}(B^*_\rho{B_\rho}+I)^{-1}=(B^*_\rho{B_\rho}+I)^{-1}\mathcal{PT}.
$$
On the other hand, $-1\in\rho(H_{-\overline{k}})$ and $\mathcal{PT}(H_k+I)^{-1}=(H_{-\overline{k}}+I)^{-1}\mathcal{PT}$ by Lemma \ref{neww14}.
Therefore
\begin{equation}\label{dede1}
\mathcal{PT}T_k=[(H_{-\overline{k}}+I)^{-1}-(H_\mu+I)^{-1}]\mathcal{PT}=T_{-\overline{k}}\mathcal{PT},
\end{equation}
where the bounded operator $T_{-\overline{k}}=(H_{-\overline{k}}+I)^{-1}-(H_\mu+I)^{-1}$ determines $H_{-\overline{k}}$ in
(\ref{sese2}).

By virtue of (\ref{dede1}), $\mathcal{PT}[I-2(1+ik)T_k]=[I-2(1+i(-\overline{k}))T_{-\overline{k}}]\mathcal{PT}$. Hence
$0\in\rho(I-2(1+i(-\overline{k}))T_{-\overline{k}})$ and the existence of ${\mathbb S}(-\overline{k})$ is
established.

Applying the operator $\mathcal{PT}$ to the both parts of (\ref{red1})
and using (\ref{dede1}) we complete the proof.
\rule{2mm}{2mm}
\subsection{Schr\"{o}dinger operator with $\mathcal{PT}$-symmetric local potential on $\mathbb{R}$.}
Denote by $\mathcal{P}f(x)=f(-x)$ and $\mathcal{T}f(x)=\overline{f(x)}$ the unitary involution and conjugation operators in
$\mathfrak{H}=L_2(\mathbb{R})$ and consider the differential expression (\ref{rada1}) with $\mathcal{PT}$-symmetric potential
$q(x)$. In that case (\ref{rada1}) determines a $\mathcal{PT}$-symmetric  $\rho$-perturbed operator
$H$.

By analogy with subsection 2.4, the image set $\{H_k\}_{k\in\Lambda_+}$ of $H$ is determined by the matrices ${\textsf T}_k=\|t_{ij}\|_{ij}^2$ in (\ref{sese2c}). Substituting the values $f_j(\pm\rho)$ and $f'_j(\pm\rho)$ of functions $f_j$ from (\ref{deder1}) into
(\ref{sese2c}) and solving the corresponding system of linear equations, we get
$$
\begin{array}{ll}
\displaystyle{t_{11}=\frac{1}{2\theta\Delta_k}[\Delta_k-e^{i\phi}(e^{i\alpha}-1)(R_k^l+e^{i(\alpha+\phi)})]}; & \displaystyle{t_{12}=\frac{T_k^l}{2\theta\Delta_k}e^{i\phi}(e^{i\alpha}-1)}; \vspace{4mm} \\
\displaystyle{t_{22}=\frac{1}{2\theta\Delta_k}[\Delta_k-e^{i\phi}(e^{i\alpha}-1)(R_k^r+e^{i(\alpha+\phi)})]}; & \displaystyle{t_{21}=\frac{T_k^r}{2\theta\Delta_k}e^{i\phi}(e^{i\alpha}-1)},
\end{array}
$$
where $\theta=1+ik$, \ $\displaystyle{e^{i\alpha}=\frac{\overline{\theta}}{\theta}}$, \ $\displaystyle{e^{i\phi}=e^{-2i\rho\textsf{Re} k}}$, \ $k\in\mathbb{C}_+'$, \ and
\begin{equation}\label{deder2}
\Delta_k=\det\left(\begin{array}{cc}
R_k^r+e^{i(\alpha+\phi)},  &  T_k^r \\
T_k^l,  & R_k^l+e^{i(\alpha+\phi)}
\end{array}\right).
\end{equation}

Obviously, if $\Delta_k\not=0$, then the entries $t_{ij}$ of ${\textsf T}_k$ are well defined. This means that the corresponding operator $T_k$
in $\mathcal{H}$ is bounded if and only if $\Delta_k\not=0$.

Further, $0\in\rho(I-2(1+ik)T_k)$ if and only if $\det(\sigma_0-2\theta{\textsf T}_k)\not=0$, where
$\sigma_0=\left(\begin{array}{cc} 1 & 0 \\
0 & 1 \end{array}\right)$. The explicit expressions of $t_{ij}$ obtained above lead to the conclusion that
$$
\sigma_0-2\theta{\textsf T}_k=\displaystyle{\frac{e^{i\phi}(e^{i\alpha}-1)}{\Delta_k}}\left(\begin{array}{cc}
R_k^l+e^{i(\alpha+\phi)},  &  -T_k^l \\
-T_k^r,  & R_k^r+e^{i(\alpha+\phi)}
\end{array}\right).
$$
Hence,
$$
\det(\sigma_0-2\theta{\textsf T}_k)=\displaystyle{\frac{e^{2i\phi}(e^{i\alpha}-1)^2}{\Delta_k}},
$$
where  $e^{i\alpha}-1\not=0$ (since $k\in\mathbb{C}_+'$). Therefore, the condition $\Delta_k\not=0$
ensures that $0\in\rho(I-2(1+ik)T_k)$.

Denote
$$
\Lambda_+'=\Lambda_+\cap\mathbb{C}_+'=\{k\in\mathbb{C}_+ : \quad \textsf{Re} \ k\not=0, \quad k^2\in\rho(H)\}.
$$
Summing up the discussion above, we establish the following
\begin{theorem}\label{new345}
The $S$-matrix ${\textsf S}(\cdot)$ of a $\mathcal{PT}$-symmetric $\rho$-perturbed operator $H$
is defined by (\ref{rest2}) for all $k\in\Lambda_+'$ such that $\Delta_k\not=0$.
\end{theorem}
\subsection{The $S$-matrix for $\mathcal{PT}$-symmetric $\rho$-perturbed operators with $\mathcal{C}$-symmetry.}
The property of $\mathcal{P}\mathcal{T}$-symmetry of $H$ is significant, but we have still to show
that $H$ can serve as an Hamiltonian for quantum mechanics.
To do so one must demonstrate that  $H$ is self-adjoint in a \emph{Hilbert space}.
In physical literature this problem is often solved
by finding a new symmetry represented by a linear operator $\mathcal{C}$, which commutes with $H$.
More precisely, suppose we can find an operator $\mathcal{C}=e^{-Q}\mathcal{P}$, where $Q$ is a bounded
self-adjoint operator in $\mathfrak{H}$  obeying the following algebraic equations:
\begin{equation}\label{usa6}
\mathcal{P}Q=-Q\mathcal{P}, \qquad \mathcal{T}Q=-Q\mathcal{T}, \qquad \mathcal{C}H=H\mathcal{C}.
\end{equation}
The first two relations in (\ref{usa6}) imply that $\mathcal{C}^2=I$ and $\mathcal{C}\mathcal{PT}=\mathcal{PT}\mathcal{C}$.

\begin{definition}\label{neww55}
{We say that a closed densely defined operator $H$ in $\mathfrak{H}$
\emph{has the property of $\mathcal{C}$-symmetry} if relations (\ref{usa6}) hold
for some choice of $\mathcal{C}=e^{-Q}\mathcal{P}$}.
\end{definition}

As a rule, if a physically meaningful $\mathcal{PT}$-symmetric operator $H$ has the
property of $\mathcal{C}$-symmetry realized by an operator $\mathcal{C}=e^{-Q}\mathcal{P}$,
then $H$ turns out to be a self-adjoint operator in the Hilbert space $\mathfrak{H}$  with
the new inner product $(e^{Q}\cdot,\cdot)$, which is equivalent to the initial inner product $(\cdot,\cdot)$.

In what follows we will use the notation $(\mathfrak{H}, (e^{Q}\cdot,\cdot))$ for the Hilbert space $\mathfrak{H}$ endowed with
inner product $(e^{Q}\cdot,\cdot)$.

Consider a  $\mathcal{PT}$-symmetric $\rho$-perturbed operator $H$ with the
property of $\mathcal{C}$-symmetry realized by an operator $\mathcal{C}=e^{-Q}\mathcal{P}$
and assume that ${\mathbb S}(\cdot)$ is the $S$-matrix of $H$ (see subsection 3.2).
It seems natural that ${\mathbb S}(\cdot)$ will contain some information about the metric operator $e^{Q}$.  We are aiming
to investigate this problem for the simplest case when the operator
$B$ has a property of $\mathcal{C}$-symmetry realized by \emph{the same} operator $\mathcal{C}=e^{-Q}{\mathcal{P}}$.
\begin{lemma}\label{newwww45}
Let $B$ satisfy (\ref{bebe3}) and let $B$ have a $\mathcal{C}$-symmetry realized by an operator $\mathcal{C}=e^{-Q}{\mathcal{P}}$.
Then $B$ keeps being a simple maximal symmetric operator in the Hilbert space
$(\mathfrak{H}, (e^{Q}\cdot,\cdot))$ and its adjoint operator in $(\mathfrak{H}, (e^{Q}\cdot,\cdot))$
coincides with the initial adjoint operator $B^*$ with respect to $(\cdot,\cdot)$.
\end{lemma}
\emph{Proof.} If $B$ has a $\mathcal{C}$-symmetry, then $B\mathcal{C}=\mathcal{C}B$, where
$\mathcal{C}=e^{-Q}\mathcal{P}$.
Then
$Be^{Q}=B{\mathcal{PC}}={\mathcal{P}}B{\mathcal{C}}={\mathcal{PC}}B=e^{Q}B$ and, hence,
$(Be^{Q})^*=e^{Q}B^*=(e^{Q}B)^*=B^*e^{Q}$. Thus, we show
\begin{equation}\label{er1}
Be^{Q}=e^{Q}B, \qquad B^*e^{Q}=e^{Q}B^*
\end{equation}
that completes the proof. \rule{2mm}{2mm}

It follows from (\ref{er1}) that $V(t)=e^{iBt}$ commutes with $e^{Q}$. Hence, the subspaces $\mathfrak{H}_\rho=V(\rho)\mathfrak{H}$ reduce
$e^{Q}$ and
\begin{equation}\label{er2}
B_{\rho}e^{Q}=e^{Q}B_\rho, \qquad B_\rho^*e^{Q}=e^{Q}B_\rho^*,
\end{equation}
where  $B_\rho$ is defined by (\ref{nee3}). By analogy with Lemma \ref{newwww45} we obtain
\begin{lemma}\label{neww45b}
Let $B$ satisfy (\ref{bebe3}) and let $B$ have a $\mathcal{C}$-symmetry realized by an operator $\mathcal{C}=e^{-Q}{\mathcal{P}}$. Then
the operator $B_\rho$ is simple maximal symmetric in both of the Hilbert spaces $(\mathfrak{H}_\rho, (\cdot,\cdot))$ and  $(\mathfrak{H}_\rho, (e^{Q}\cdot,\cdot))$ and its adjoint $B_\rho^*$ does not depend on the choice of inner products $(\cdot,\cdot)$ or $(e^{Q}\cdot,\cdot)$.
\end{lemma}
\begin{theorem}\label{neww78}
Let a  $\mathcal{PT}$-symmetric $\rho$-perturbed operator ${H}$ have nonnegative real spectrum with $0\not\in\sigma_p(H)$
and let $H$ be self-adjoint in the Hilbert space $(\mathfrak{H}, (e^{Q}\cdot,\cdot))$, where $\mathcal{C}=e^{-Q}{\mathcal{P}}$ is an operator
of  $\mathcal{C}$-symmetry of a simple maximal symmetric operator $B$ satisfying (\ref{bebe3}). Then
the $S$-matrix ${\mathbb S}(k)$ of $H$ is an analytic operator-valued function in $\mathbb{C}_+$, which is determined by (\ref{red1}), and for all $k\in\mathbb{C}_+$:
\begin{equation}\label{neww69}
\begin{array}{c}
(i) \quad  e^{Q}{\mathbb S}(-\overline{k})={\mathbb S}^*(k)e^{Q}; \vspace{3mm} \\
(ii) \quad e^{Q}{\mathbb S}(-\overline{k}){\mathbb S}(k)\leq{e^{Q}} \vspace{3mm} \\
(iii) \quad \mathcal{PT}{\mathbb S}(-\overline{k})={\mathbb S}(k)\mathcal{PT}.
\end{array}
\end{equation}
\end{theorem}
\emph{Proof.} Let us consider the operators $H_\mu=B^*B$ and $H$ in the Hilbert space $(\mathfrak{H}, (e^{Q}\cdot,\cdot))$.
 Due to Lemma \ref{newwww45}, the operator $H_\mu$ does not depend on the choice of $\mathcal{C}$-symmetry of $B$ and $H_\mu$ is an unperturbed operator in $(\mathfrak{H}, (e^{Q}\cdot,\cdot))$ in the sense of Definition \ref{did1}.

On the other hand, Lemma \ref{neww45b} and Definition \ref{did2} imply that $H$ is a $\rho$-perturbed self-adjoint operator
in $(\mathfrak{H}, (e^{Q}\cdot,\cdot))$ with nonnegative spectrum. Moreover, the condition $0\not\in\sigma_p(H)$ yields that
$H$ is a positive self-adjoint operator with respect to $(e^{Q}\cdot,\cdot)$.
This means that for the operator $H$ considered in $(\mathfrak{H}, (e^{Q}\cdot,\cdot))$ there exists
the $S$-matrix ${\mathbb S}(k)$ in the spectral representation $L_2({\mathbb R}_+,{\mathcal N})$ of $H_\mu$
(see subsection 2.2).

The $S$-matrix ${\mathbb S}(k)$ is defined by formula (\ref{red1}) for all $k\in\mathbb{C}_+$ and its
calculation consists of three stages, which have to be done in $(\mathfrak{H}, (e^{Q}\cdot,\cdot))$:
\begin{itemize}
  \item[(A)] the determination of the image set $\{H_k\}_{k\in\mathbb{C}_+}$ of $H$;
  \item[(B)] the construction of the positive boundary triplet $(\mathcal{H}, \Gamma_0, \Gamma_1)$;
  \item[(C)] the determination of operators $\{T_k\}$ in (\ref{sese2}).
\end{itemize}

Let us to examine the dependence of these stages on the change of inner product: $(e^{Q}\cdot,\cdot)\to(\cdot,\cdot)$.

 (A) \emph{The image set.}
Since ${H}$ is a positive self-adjoint $\rho$-perturbed operator in $(\mathfrak{H}, (e^{Q}\cdot,\cdot))$, we can define the image set
$\{H_k\}_{k\in\mathbb{C}_+}$ of $H$ by the formula (\ref{ada52}), where $P_\rho$ is the orthogonal projection onto the subspace $\mathfrak{H}_\rho$ in $(\mathfrak{H}, (e^{Q}\cdot,\cdot))$.
Since the subspace $\mathfrak{H}_\rho$ reduces $e^{Q}$,
the orthogonal decomposition $\mathfrak{H}=\mathfrak{H}_\rho\oplus\kerr{V^*(\rho)}$ with respect to $(e^{Q}\cdot,\cdot)$ remains orthogonal for inner product $(\cdot,\cdot)$. This means that $P_\rho$ is also the orthogonal projection onto $\mathfrak{H}_\rho$  in $\mathfrak{H}$ with respect to the inner product $(\cdot,\cdot)$. Therefore, the image set $\{H_k\}_{k\in\mathbb{C}_+}$ determined by (\ref{ada52})
\emph{does not depend on the choice of inner product} $(e^{Q}\cdot,\cdot)$ or $(\cdot,\cdot)$.

(B) \emph{The positive boundary triplet $(\mathcal{H}, \Gamma_0, \Gamma_1)$.}  Lemma  \ref{neww45b} and relations (\ref{er2}) imply that:  the subspace $\mathcal{H}=\kerr({B^*_\rho}^2+I)$ does not depend on the choice $(e^{Q}\cdot,\cdot)$ or $(\cdot,\cdot)$; \  the subspace $\mathcal{H}$ reduces $e^{Q}$; \ the operator ${B^*_\rho}{B_\rho}$ commutes with $e^{Q}$. Therefore, the decomposition (\ref{bonn41}) and the operator $\Gamma_0$ in (\ref{e7}) do not depend on the choice of inner product.

Furthermore, since $\mathcal{H}$ reduces $e^{Q}$, the
orthogonal decomposition $\mathfrak{H}_\rho=\mathcal{H}\oplus\mathcal{R}(B^2_\rho+I)$ does not depend on the choice of inner product
$(e^{Q}\cdot,\cdot)$ or $(\cdot,\cdot)$. Hence, the orthogonal projection $P_{\mathcal{H}}$ does not change and the
operator $\Gamma_1$ in (\ref{e7}) does not depend on the choice of inner products. Thus, we show that
the formulas (\ref{bonn41}) and (\ref{e7}) determine the boundary triplet
$(\mathcal{H}, \Gamma_0, \Gamma_1)$ of $B_\rho^2$, which \emph{does not depend on the choice of inner product} $(e^{Q}\cdot,\cdot)$ or $(\cdot,\cdot)$.

(C) \emph{Operators $\{T_k\}$.} It follows from (A) and (B) that the operators $\{T_k\}$ describing the image set $\{H_k\}_{k\in\mathbb{C}_+}$ in (\ref{sese2}) do not depend on the choice of inner product $(e^{Q}\cdot,\cdot)$ or $(\cdot,\cdot)$.

Combining (A) - (C) we arrive at the conclusion that the $S$-matrix ${\mathbb S}(\cdot)$ defined by (\ref{red1}) \emph{does not depend on the choice of $(e^{Q}\cdot,\cdot)$ or $(\cdot,\cdot)$}.

Due to Theorem \ref{esse3} and Remark \ref{esse3d}, the $S$-matrix
${\mathbb S}(\cdot)$ is an analytic operator-valued function in $\mathbb{C}_+$
and
\begin{equation}\label{neww1}
{\mathbb S}(-\overline{k})={\mathbb S}^{[*]}(k), \qquad \forall{k}\in\mathbb{C}_+,
\end{equation}
where ${\mathbb S}^{[*]}$ is the adjoint operator in $\mathcal{H}$ with respect to $(e^{Q}\cdot,\cdot)$.
It is clear that
\begin{equation}\label{neww31}
e^{Q}{\mathbb S}^{[*]}(k)={\mathbb S}^{*}(k)e^{Q},
\end{equation}
where ${\mathbb S}^{*}$ is the adjoint operator in $\mathcal{H}$ with respect to $(\cdot,\cdot)$.

The relation (i) in (\ref{neww69}) follows from (\ref{neww1}) and (\ref{neww31}).

Further, the operators ${\mathbb S}(k)$ are contraction operators
with respect to $(e^{Q}\cdot,\cdot)$ in $\mathcal{H}$. Hence
$$
(e^{Q}{\mathbb S}(k)f,{\mathbb S}(k)f)=(e^{Q}{\mathbb S}^{[*]}(k){\mathbb S}(k)f,f)\leq(e^{Q}f,f), \qquad \forall{f}\in\mathcal{H}.
$$
The obtained inequality and (\ref{neww1}) mean that
$$
e^{Q}{\mathbb S}(-\overline{k}){\mathbb S}(k)=e^{Q}{\mathbb S}^{[*]}(k){\mathbb S}(k)\leq{e^{Q}}
$$
with respect to the inner product $(\cdot,\cdot)$. Thus the relation (ii) in (\ref{neww69})
is established. Relation (iii) follows from Proposition \ref{neww28}.
Theorem \ref{neww78} is proved. \rule{2mm}{2mm}

\smallskip

Since the operators $\mathcal{P}$, $\mathcal{PT}$, and $Q$ are reduced by the subspace $\mathcal{H}=\kerr({B_\rho^*}^2+I)$, the formulas
\begin{equation}\label{neww24}
\textsf{P}=Y\mathcal{P}Y^{-1}, \quad \textsf{T}=Y\mathcal{T}Y^{-1}, \quad \textsf{Q}=YQY^{-1},
\end{equation}
where $Y$ isometrically maps $\mathcal{H}$ onto the subspace
$\{e^{-\delta}u : {u}\in{\mathcal N}\}$ of ${L_2((\rho,+\infty),{\mathcal N})}$ with the subsequent identification (\ref{rest1}),
determine, respectively, a unitary involution $\textsf{P}$, a conjugation $\textsf{T}$, and a self-adjoint operator $\textsf{Q}$ in ${\mathcal N}$. Considering the operator-valued function
$$
{\textsf S}(k)=Y{\mathbb S}(k)Y^{-1}
$$
with values in ${\mathcal N}$ and taking (\ref{neww24}) into account we rewrite Theorem \ref{neww78} as follows.
\begin{theorem}\label{neww144}
Under conditions of Theorem \ref{neww78}, the $S$-matrix ${\textsf S}(k)$  is an analytic operator-valued function in $\mathbb{C}_+$, which is determined by (\ref{red1b}), and has the following properties:
\begin{equation}\label{neww69b}
\begin{array}{c}
(i) \quad  e^{\textsf{Q}}{\textsf{S}}(-\overline{k})={\textsf{S}}^*(k)e^{\textsf{Q}}; \vspace{3mm} \\
(ii) \quad e^{\textsf{Q}}{\textsf{S}}(-\overline{k}){\textsf{S}}(k)\leq{e^{\textsf{Q}}}; \vspace{3mm} \\
(iii) \quad \textsf{PT}{\textsf S}(-\overline{k})={\textsf S}(k)\textsf{PT},
\end{array}
\end{equation}
where ${\textsf{S}}^*$ is the adjoint operator in $\mathcal{N}$.
\end{theorem}

\subsection{An example of inverse problem solution.} Relations (\ref{neww69b}) contain information about the metric operator
$e^Q$ and the corresponding inner product $(e^Q\cdot,\cdot)$ which ensures the self-adjointness of $H$.
This leads to natural question: \emph{is it possible for a given $\mathcal{PT}$-symmetric operator $H$ to recover the corresponding
metric operator $e^Q$ by the $S$-matrix ${\textsf{S}}(\cdot)$?}

A simple example considered below show that the answer is positive for certain classes of  $\mathcal{PT}$-symmetric operators.

Consider the one-dimensional Schr\"{o}dinger differential expression with singular zero-range potential
\begin{equation}\label{as3}
-\frac{d^2}{dx^2}+q_\gamma(x), \quad
q_\gamma(x)=i\gamma(<\delta',\cdot>\delta(x)+
 <\delta,\cdot>\delta'(x)), \ \gamma\in\mathbb{R},
\end{equation}
where $\delta(x)$ and $\delta'(x)$ are, respectively, the Dirac
$\delta$-function and its derivative (with support at $0$).

It is easy to verify that $\mathcal{PT}q_\gamma(x)=q_\gamma(x)\mathcal{PT}$,
where $\mathcal{P}$ is the space parity operator $\mathcal{P}f(x)={f(-x)}$ and $\mathcal{T}$ is the complex conjugation operator
$\mathcal{T}f(x)=\overline{f(x)}$.

The operator realization $H_\gamma$ of (\ref{as3}) in $L_2(\mathbb{R})$ is defined as
$$
H_\gamma=l_{\mathrm{reg}}\upharpoonright\mathcal{D}(H_\gamma), \quad
\mathcal{D}(H_\gamma)=\{f\in{{W_2^2}(\mathbb{R}\backslash\{0\})} \ :
\ l_{\mathrm{reg}}(f)\in{L_2(\mathbb{R})}\},
$$
where the regularization $l_{\mathrm{reg}}$ of the differential expression  (\ref{as3}) onto
${W_2^2}(\mathbb{R}\backslash\{0\})$ has the form
$$
{l}_{\mathrm{reg}}(\cdot)=-\frac{d^2}{dx^2}+i\gamma(<\delta_{\mathrm{ex}}',\cdot>\delta(x)+
<\delta_{\mathrm{ex}},\cdot>\delta'(x)).
$$
Here $-{d^2}/{dx^2}$ acts on ${W_2^2}(\mathbb{R}\backslash\{0\})$ in
the distributional sense and
$$
 <\delta_{\mathrm{ex}}, f>=\frac{f(+0)+f(-0)}{2}, \quad <\delta_{\mathrm{ex}}', f>=-\frac{f'(+0)+f'(-0)}{2}
$$
for all $f(x)\in{{W_2^2}(\mathbb{R}\backslash\{0\})}$.

An equivalent description of $H_\gamma$ is (see \cite[Theorem 1]{AK}):
\begin{equation}\label{neww74}
H_\gamma=-\frac{d^2}{dx^2}, \quad \mathcal{D}(H_\gamma)=\left\{f\in{{W_2^2}(\mathbb{R}\backslash\{0\})} \ : \
\begin{array}{cc}
f(0+)=e^{i\beta}f(0-)  \\
f'(0+)=e^{-i\beta}f'(0-)
\end{array} \right\},
\end{equation}
where $e^{i\beta}=\frac{2+i\gamma}{2-i\gamma}$.

The operator $H_\gamma$ is $\mathcal{PT}$-symmetric in $L_2(\mathbb{R})$, its spectrum coincides with $\mathbb{R}_+$ for $|\gamma|\not=2$ and
$\sigma(H_\gamma)=\mathbb{C}$ for $|\gamma|=2$ (\cite[Theorem 2]{AK}).

It follows from (\ref{ses1}) and (\ref{neww74}) that $H_\gamma$ is $0$-perturbed. Therefore, for $|\gamma|\not=2$, we can determine
the $S$-matrix ${\textsf{S}}(\cdot)$ by the formula (\ref{rest2}).

The coefficients $T_k^l, T_k^r, R_k^l, R_k^r$ in (\ref{rest2})
are determined by the condition that functions $f_j$ in (\ref{deder1}) belong to $\mathcal{D}(H_\gamma)$ in (\ref{neww74}).
Simple calculations give
$$
T_k^l=T_k^r={\frac{\textsf{Re}\ k}{k\cos\beta}}, \quad R_k^r={i\frac{(\textsf{Re}\ k)\sin\beta-(\textsf{Im}\ k)\cos\beta}{k\cos\beta}},
$$
and $\displaystyle{R_k^l=-{i\frac{(\textsf{Re}\ k)\sin\beta+(\textsf{Im}\ k)\cos\beta}{k\cos\beta}}}$.
Substituting these quantities into (\ref{rest2}) and taking into account that $\rho=0$, we obtain
\begin{equation}\label{deder7}
{\textsf S}(k)=-\left(\begin{array}{cc}
i\tan\beta & \frac{1}{\cos\beta}  \vspace{4mm} \\
\frac{1}{\cos\beta} & -i\tan\beta
\end{array}
\right).
\end{equation}
Thus, the $S$-matrix ${\textsf S}(k)$ is a constant in $\mathbb{C_+}$ and it is defined for all $|\gamma|\not=2$.
If $|\gamma|=2$, i.e., $\gamma=2$ or $\gamma=-2$, then $\beta=\frac{\pi}{2}$ or $\beta=-\frac{\pi}{2}$ and
the  $S$-matrix does not exist in  $k\in\mathbb{C}_+$ (the entries of ${\textsf S}(k)$ are $\infty$).
This is natural because $\sigma(H_{\pm{2}})=\mathbb{C}$.

In our case, the isometric mapping $Y:\mathcal{H}\to\mathbb{C}^2$ is defined by (\ref{deder8}). Then the image $\textsf{P}$
of $\mathcal{P}$ in (\ref{neww24}) has the form $\textsf{P}=\sigma_1=\left(\begin{array}{cc}
0 & 1 \\
1 & 0 \end{array}
\right)$. The image $\textsf{T}$ of $\mathcal{T}$ is the standard operator of conjugation in $\mathbb{C}^2$.

For such choice of $\textsf{P}$ and $\textsf{T}$ the relation (iii) of (\ref{neww69b}) holds for the $S$-matrix (\ref{deder7}).
Let us suppose that the relation (i) holds for some choice of self-adjoint operator $\textsf{Q}$ in $\mathbb{C}^2$, i.e.
\begin{equation}\label{deder9}
e^{\textsf{Q}}\left(\begin{array}{cc}
i\tan\beta & \frac{1}{\cos\beta}  \vspace{4mm} \\
\frac{1}{\cos\beta} & -i\tan\beta
\end{array}
\right)=\left(\begin{array}{cc}
-i\tan\beta & \frac{1}{\cos\beta}  \vspace{4mm} \\
\frac{1}{\cos\beta} & i\tan\beta
\end{array}
\right)e^{\textsf{Q}}.
\end{equation}

The operator $\textsf{Q}$ is a $2\times{2}$ matrix. Hence,
\begin{equation}\label{deder10}
\textsf{Q}=\chi_0\sigma_0+\chi_1\sigma_1+\chi_2\sigma_2+\chi_3\sigma_3, \qquad \chi_j\in\mathbb{C},
\end{equation}
where $\sigma_3=\left(\begin{array}{cc}
1 & 0 \\
0 & -1 \end{array}
\right)$ and $\sigma_2=i\sigma_1\sigma_3=\left(\begin{array}{cc}
0 & -i \\
i & 0 \end{array}
\right)$ are the Pauli matrices.

It follows from (\ref{usa6}) and (\ref{neww24}) that
$$
\textsf{Q}=\textsf{Q}^*, \quad \textsf{PQ}=-\textsf{QP}, \quad \textsf{TQ}=-\textsf{QT}.
$$
Imposing these conditions in (\ref{deder10}) we derive that $\textsf{Q}={\chi\sigma_2}$, where $\chi\in\mathbb{R}$.
Hence, $e^\textsf{Q}=e^{\chi\sigma_2}=(\cosh\chi)\sigma_0+(\sinh\chi)\sigma_2$.
Substituting this expression into (\ref{deder9}) and making elementary calculations we arrive at the conclusion that
the relation (\ref{deder9}) is true if and only if
\begin{equation}\label{deder11}
\tanh\chi=\sin\beta.
\end{equation}
Thus, we determine uniquely the operator $e^\textsf{Q}=e^{\chi\sigma_2}$ which satisfies relation (i)
in (\ref{neww69b}) with the $S$-matrix (\ref{deder7}).

Denote by $\mathcal{R}$ the operator of unitary involution $\mathcal{R}f(x)=(\textsf{sgn}\ x)f(x)$
in $L_2(\mathbb{R})$. The subspace $\mathcal{H}$ reduces $\mathcal{R}$ and its image $\textsf{R}=Y\mathcal{R}Y^{-1}$
in $\mathbb{C}^2$ coincides with $\sigma_3$. Therefore, the image of unitary involution $i\mathcal{PT}$ in $\mathbb{C}^2$
will coincide with $i\textsf{PR}=i\sigma_1\sigma_3=\sigma_2$.

Taking this relationship and Theorems \ref{neww78}, \ref{neww144} into account, it  is natural to suppose that the operator $H_\gamma$ will have the property of $\mathcal{C}$-symmetry realized by the operator
\begin{equation}\label{deder12}
\mathcal{C}=e^{-i\chi\mathcal{PR}}\mathcal{P}=(\cosh\chi)\mathcal{P}+i(\sinh\chi)\mathcal{R},
\end{equation}
where $\chi$ is determined by (\ref{deder11}) and $H_\gamma$ turns out to
be a self-adjoint operator with respect to new inner product
$(e^{i\chi\mathcal{PR}}\cdot,\cdot)$ in $L_2(\mathbb{R})$.

This assumption is true due to the results \cite[Subsection 5.1]{KKK}, where was shown that the operator $H_\gamma$ has the $\mathcal{C}$-symmetry (\ref{deder12}) and $H_\gamma$ is a self-adjoint operator with respect to $(e^{i\chi\mathcal{PR}}\cdot,\cdot)$.

\section{Summary and Discussion}
In the present paper an operator-theoretical interpretation of the Lax-Phillips scattering theory \cite{LF} developed in
\cite{KU1,AlAn,Kioto} has been reshaped to define and calculate $S$-matrices
of $\mathcal{PT}$-symmetric $\rho$-perturbed operators.
The crucial role is played by the concept of $\rho$-perturbed operator (Definition \ref{did2})
that allows one to consider the scattering of many concrete systems with perturbation supported
at bounded domain from a unique point of view.

In our opinion, the advantages of the proposed approach have their origin in the
following properties:
\begin{enumerate}
  \item[(i)] for typical examples, the general formula (\ref{red1}) for $S$-matrix
leads to well known expressions derived by the standard methods (see subsection 3.3 or \cite[Section 3]{AlAn});
  \item[(ii)] the  formula (\ref{red1})  ensure simple links to powerful mathematical methods of the extension
  theory of symmetric operators.
\end{enumerate}

We believe that interplay between (i) and (ii) will provide some deeper insights into the structural subtleties of
inverse scattering problems for $\mathcal{PT}$-symmetric operators.

Let us illustrate this point by considering the case
of positive self-adjoint $\rho$-perturbed operators. In that case the corresponding $S$-matrix (\ref{red1}) uniquely determines
the image set $\{H_k\}_{k\in\mathbb{C}_+}$ by formula (\ref{sese2}). The image set $\{H_k\}_{k\in\mathbb{C}_+}$
determines a generalized resolvent of the symmetric operator $B^2_\rho$ \cite{AkG1,Sh}:
$$
R_k=(H_k-k^2)^{-1}, \qquad k\in\mathbb{C}_+.
$$
This means that
$$
(R_kf,g)=\int_{-\infty}^\infty\frac{d(F_tf,g)}{t-k}, \qquad {f,g}\in\mathfrak{H}_\rho,
$$
where $F_t$ is a spectral function of the symmetric operator  $B^2_\rho$. By the Naimark theorem \cite{Naimark}, the spectral function $F_t$
uniquely determines a minimal positive self-adjoint $\rho$-perturbed operator $H$ up to unitary equivalence (see \cite[Section 4]{AlAn} for detail). Thus, we derive the solution of inverse problem in the following form:

$S$-matrix $\to$ image set $\{H_k\}_{k\in\mathbb{C}_+}$  $\to$ generalized resolvent $R_k$ $\to$ spectral function $F_t$ of symmetric operator
$B^2_\rho$ $\to$ $\rho$-perturbed operator $H$.

It seems highly interesting to extend at least part of these conclusions to the case of $\mathcal{PT}$-symmetric $\rho$-perturbed operators $H$
with $\mathcal{C}$-symmetries $\mathcal{C}=e^{-Q}\mathcal{P}$ trying to recover an information about the metric operator $e^{Q}$
from the $S$-matrix.

The methods developed in the present paper can also be applied to Dirac systems with $\rho$-perturbation type.
We hope to undertake those discussions in a forthcoming paper.

\smallskip

\noindent \textbf{Acknowledgements.}
Research supported by the Polish Ministry of Science and Higher Education: 11.11.420.04 and N201546438.
The second named author thanks JRP IZ73Z0 (28135) of SCOPES 2009-2012
for support.

\end{document}